\documentstyle[epsfig]{jaa}

\DeclareMathAlphabet{\mathsc}{OT1}{cmr}{m}{sc}
\def\testbx{bx}%
\DeclareRobustCommand{\ion}[2]{%
\relax\ifmmode
\ifx\testbx\f@series
{\mathbf{#1\,\mathsc{#2}}}\else
{\mathrm{#1\,\mathsc{#2}}}\fi
\else\textup{#1\,{\mdseries\textsc{#2}}}%
\fi}
\def\ch{\footnotesize}

\def\aa{Astron. Astrophys.}
\def\ApJ{Astrophys. J.}
\def\MNRAS{Mon. Not. R. Astron. Soc.}

\begin{document}
\title[A hidden radio halo in A\,1682?]
{A hidden radio halo in the galaxy cluster A\,1682?}
\author[T. Venturi et al.] 
{T. Venturi,$^1$\thanks{e-mail: tventuri@ira.inaf.it}
S. Giacintucci,$^2$, D. Dallacasa$^{3,1}$\\
%\newauthor
%R. Athreya$^5$, G. Brunetti,$^1$ R. Cassano,$^1$ G. Macario,$^1$\\
$^1$INAF, Istituto di Radioastronomia, Via Gobetti 101, 40129 Bologna, Italy\\
$^2$Department of Astronomy, University of Maryland, College Park, MD 20742--2421, USA\\
$^3$ Department of Astronomy, University of Bologna, Via Ranzani 1, 40129 Bologna, Italy}
%$^4$ Naval Research Laboratory, Code 7213, Washington, DC 20375, USA\\
%$^5$ Indian Institute of Science Education \& Research, Pune, India}
\pubyear{xxxx}
\volume{xx}
\date{Received xxx; accepted xxx}
\maketitle
\label{firstpage}
\begin{abstract}

High sensitivity observations of radio halos in galaxy clusters at 
frequencies $\nu\le 330$ MHz are still relatively rare, and very
little is known compared to the classical 1.4 GHz images.
The few radio halos imaged down to 150--240 MHz show a considerable spread 
in size, morphology and spectral properties. 
\\
All clusters belonging to the GMRT Radio Halo Survey with detected or
candidate cluster--scale diffuse emission have been imaged 
at 325 MHz with the GMRT. Few of them were also observed with the GMRT at
240 MHz and 150 MHz. For A\,1682, imaging is particularly challenging due 
to the presence of strong and extended radio galaxies at the center. 
Our data analysis suggests that the radio galaxies are superposed to very 
low surface brightness radio emission extended on the cluster scale, which 
we present here.

\end{abstract} 

\section{Introduction}

The origin of radio halos and relics is strictly connected to the 
dynamics of the hosting cluster.
Recent works, based on the GMRT Radio Halo Survey (Venturi et al. 
2007 \& 2008), have highlighted a number of relevant statistical 
properties of radio halos, such as the {\it bimodality}: 
clusters are either {\it radio loud},
in which case the halo radio power correlates with the cluster X--ray
luminosity, following the well--known logL$_{\rm X}$--logP$_{\rm 1.4~GHz}$
correlation, or {\it radio quiet}, with radio power upper limits well 
below the same correlation  (Brunetti et al. 2009).
Moreover, a quantitative
radio/X--ray analysis of the radio halo/cluster merger connection 
shows that halos are found only in cluster mergers; 
relaxed clusters never host them, and only few massive, 
luminous and merging clusters are void of diffuse emission at the sensitivity
level of the current radio interferometers (Cassano et al. 2010) 

These results support  the idea that radio halos are the result of 
relativistic particle re--acceleration by turbulence induced in the 
cluster volume by massive cluster mergers (the {\it re--acceleration model}, 
Petrosian 2001 and Brunetti et al. 2001).
This model further predicts the existence of a population of halos with 
``ultra--steep spectrum'', i.e. $\alpha \ge 1.5$, which would be connected 
to the  less energetic, but much more frequent minor mergers (Cassano 2009 
and Cassano, these proceedings). A\,521, with a spectral slope 
$\alpha \sim 1.9$, is the first ultra--steep spectrum radio halo (USSRH) 
discovered (Brunetti et al. 2008, Dallacasa et al. 2009).
Due to their spectral properties, USSRH are best detectable at low
frequencies, i.e. $\nu\le$330 MHz.

To improve the knowledge of the low--frequency properties of radio halos, 
and possibly identify candidate USSRHs, all clusters in the GMRT radio halo 
survey with detected diffuse or candidate cluster--scale emission were 
observed with the GMRT at 325 MHz, and some of them were followed up
at 240 and 150 MHz. 
Some results have already been published in recent papers  
(Giacintucci et al. 2009 \& 2011, Macario et al. 2010 \& 2011, 
Venturi et al. 2011). Here we present and briefly 
discuss the challenging case of A\,1682. 

\section{A hidden radio halo in A\,1682?}

A\,1682 (z=0.2260) 
%(RA$_{\rm J2000}={\rm 13^h06^m49.07^s}$, 
%DEC$_{\rm J2000}=+46^{\circ}32^{\prime}59^{\prime\prime}$, 
is a merging massive cluster (Morrison et al. 2003), with 
L$_{\rm X}{\rm [0.2-2.4keV]}=7.02\times10^{44}$ erg s$^{-1}$, first imaged
at high sensitivity and resolution with the GMRT at 610 MHz 
(Venturi et al. 2008). The radio emission is dominated by a strong 
tailed radio galaxy at the cluster center, and by two features of unclear 
nature, referred to as S--E and N--W ridge (see Fig. 6 in Venturi et al. 2008). 
Beyond those individual sources, residuals are present in the central 
cluster region, suggestive of a radio halo outshined by the bright individual 
radio sources. We carried out follow--up observations of A\,1682 over a 
wide frequency range, in 
order to study the spectral properties of the individual sources, 
and possibly image the underlying residual cluster--scale emission. 

\begin{table}[h]
\caption[]{A\,1682 observing logs and information on the full resolution images}
\begin{center}
\begin{tabular}{cccccc}
\hline
\ch {\bf $\nu$}& \ch {\bf Array}& \ch {\bf Date} & Time & HPBW & rms \\
MHz            &                &                &  hr  & 
$^{\prime\prime}\times^{\prime\prime}$,$^{\circ}$ & mJy b$^{\rm -1}$ \\
\hline
610 & GMRT     & 05-12-2006 & 6  & 6.2$\times4.1$, 61.2 & 0.03\\
330 & GMRT     & 06-02-2011 & 16 & 10.0$\times9.1$, --  & --  \\
240 & GMRT     & 05-12-2006 & 6  & 12.5$\times9.2$,55.7 & 0.50\\
150 & GMRT     & 17-08-2009 & 9  & 26.0$\times19.0$, -- & --  \\
~74 & VLA--A   & 08-01-2009 & 4  & 23.8$\times22.4$,--68 & 20 \\
\hline
\end{tabular} 
\end{center}
\label{tab:obs} 
\end{table}

Table \ref{tab:obs} lists the observations and the available information.
We completed the data imaging and analysis at 610 MHz and 240 MHz and
at 74 MHz. The data reduction at 330 MHz and 150 MHz is 
currently in progress. At 240 and 610 MHz,
after removal of the RFI, editing and self--calibration, we produced 
full resolution images with a 1--k$\lambda$ gap in the u--v plane in 
order to image the individual sources without the inclusion of the 
cluster  scale diffuse emission. We subtracted the clean components 
from the GMRT datasets, and imaged the residuals using the whole u--v range
and a resolution $\sim 35^{\prime\prime}$--$40^{\prime\prime}$.
The rms in the residual images are $\sim$ 0.12 mJy/b and 0.8 mJy/b 
at 610 MHz and 240 MHz respectively.
The 74 MHz data reduction was carried out applying the model of Cygnus A 
provided by the NRAO for the bandpass and amplitude calibation. Then 
editing and self--calibration were performed. 
\\
Our results are shown in Fig. 1. The left panel shows the 610 MHz residual 
and full resolution image (black and red contours respectively); the right 
panel shows the 240 MHz residual image overlaid on the VLA--A 74 MHz cluster 
emission. Excess emission spread over the cluster scale 
is clearly present at both frequencies. Re--analysis of archival 
VLA D--array data at 1.4 GHz also suggests hints of residual emission.
The residual flux density values are S$_{\rm 1.4~GHz}\sim$14.5 mJy, 
S$_{\rm 610~MHz}\sim$32 mJy, S$_{\rm 240~MHz}\sim$80 mJy, 
consistent with a spectral slope of $\alpha \sim 1$. 
Considering the uncertainties in the source subtraction procedure and the
different u--v coverage at the various frequencies, any detailed consideration
on this emission and its nature is premature at this stage.
The important
piece of information is that all our images show excess emission at the 
center of A\,1682, most likely spread over the
Mpc cluster scale (1$^{\prime\prime}$=3.626 kpc at the distance of A\,1682).
At 74 MHz it is impossible to separate the contribution of the individual 
sources from the diffuse emission, however the colour scale image in the
right panel of Fig. 1  clearly shows the presence of a blob, located
S--E of the extended central radio galaxies, coincident with a 
similar feature in the 240 MHz residuals. Actually, 
a large fraction of the diffuse flux density both at 610 MHz and at 240 
MHz comes from this region.
\\
The nature of the S--E and N--W ridges is unclear. Their spectrum is steep,
$\alpha_{\rm 240~MHz}^{\rm 610~MHz}\sim$1.3 and 1.6 respectively.
The brightness distribution of the N--W ridge (centrally peaked and symmetric 
in both directions) is suggestive of a FR--I radio galaxy rather than a relic.

We expect that the GMRT 150 MHz and 330 MHz will strengthen the
suggestion of the hidden radio halo in A\,1682, and provide important 
information on its spectral behaviour.

\begin{figure}
\centering
\includegraphics[width=5.2cm, angle=0]{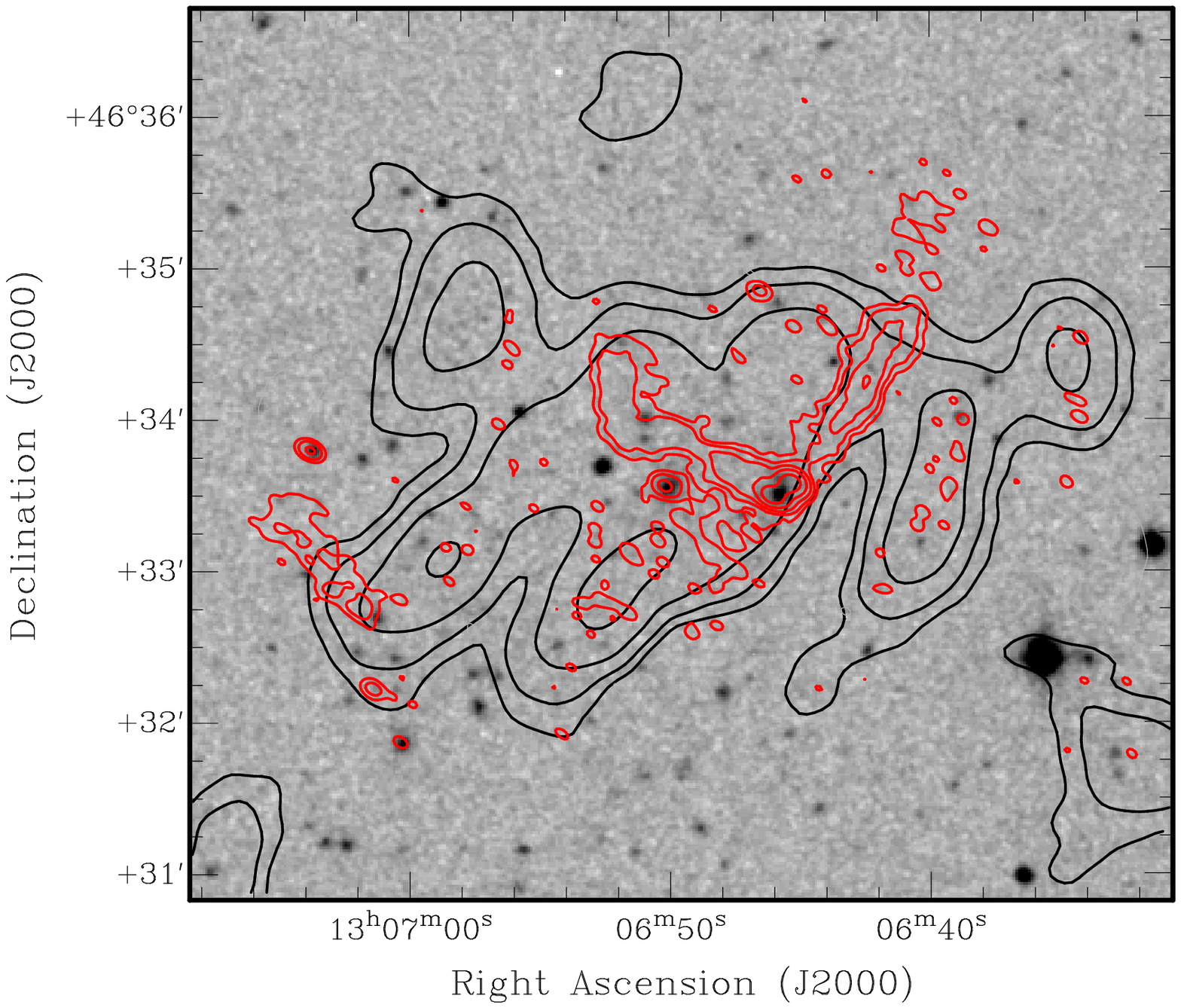}
\includegraphics[width=5.0cm, angle=0]{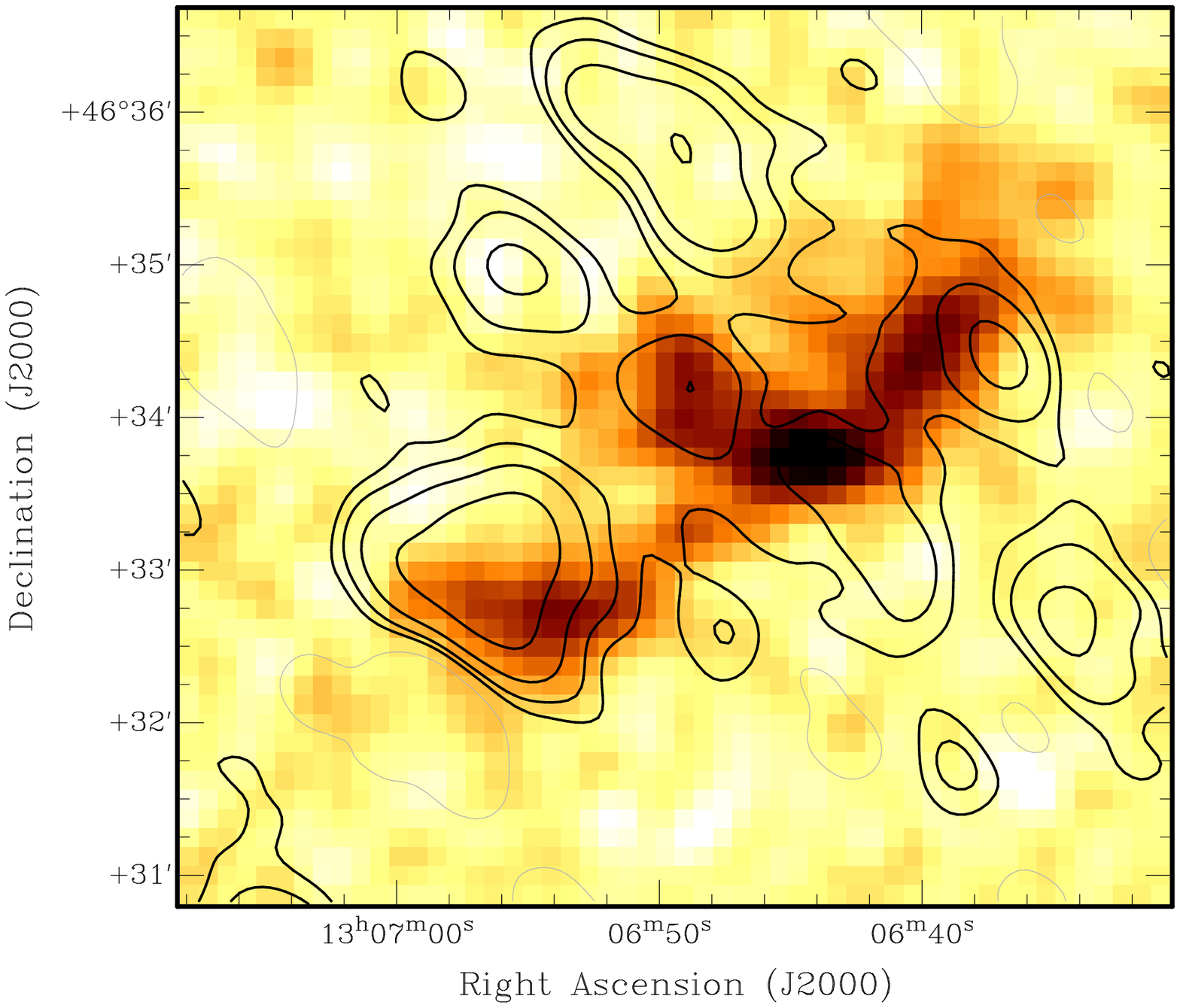}
\caption{Left -- 610 MHz GMRT contours on the optical DSS--2
frame. Red: full resolution image
(6.2$^{\prime\prime}\times4.1^{\prime\prime}$, 
contours $\pm$0.15,0.6,2.4 mJy/b). Black: residual image 
($38^{\prime\prime}\times33^{\prime\prime}$, contours 
$\pm$0.3, 0.6, 1.2, 2.4 mJy/b, $1\sigma\sim0.1$).
Right -- 240 MHz GMRT residual image ($40^{\prime\prime}\times30^{\prime\prime}$, 
black contours $\pm$ 0.6, 1.2, 2.4, 4.8 mJy/b, $1\sigma\sim0.3$ mJy/b),
overlaid on the 74 MHz VLA--A total intensity image 
($23.8^{\prime\prime}\times22.4^{\prime\prime}$, rms $\sim$20 mJy/b). }
\label{fig:images}
\end{figure}

\section*{Acknowledgments} 
We acknowledge our collaborators: G. Brunetti, R. Cassano, G. Macario, 
R. Athreya, T.E. Clarke, W.M. Lane, A. Cohen, N. Kassim and W. Cotton.
We thank L. Rudnick for the insightful discussions.
The GMRT is run by the National Center for Radio Astrophysics of the
Tata Institute of Fundamental Research. 
%The National Radio Astronomy %observatory (NRAO) 
%The NRAO is a facility of the National Science Foundation.
The NRAO %National Radio Astronomy Observatory 
is a facility of The National 
Science Foundation operated under cooperative agreement by Associated 
Universities, Inc.

\end{document}